\documentclass[twocolumn,preprintnumbers,amsmath,amssymb,prb]{revtex4} 

\usepackage[dvips]{graphicx}
\usepackage{dcolumn}
\usepackage{bm}

\begin{document}

\maketitle

%
%

{\Large\bf\noindent A quantum fluid of metallic \\ hydrogen suggested by \\ first-principles calculations}\\[0.5cm]
{\bf Stanimir A. Bonev\cite{presadr}, Eric Schwegler, \\ Tadashi Ogitsu  \& Giulia Galli} \\[0.5cm]
{\it Lawrence Livermore National Laboratory,
              University of California, Livermore, California 94550}\\[0.5cm]
{\small \bf \noindent 
It is generally assumed\cite{4,5,mao} that solid hydrogen 
will transform into a metallic alkali-like crystal at sufficiently high 
pressure.
However, some theoretical models\cite{6,10} have also suggested that 
compressed hydrogen may form an unusual two-component (protons and electrons)
metallic fluid at low temperature, or possibly even a zero-temperature liquid 
ground state.
The existence  of these new states of matter is conditional on 
the presence of a maximum in the melting temperature versus pressure curve 
(the 'melt line'). Previous measurements\cite{1,2,3} of the 
hydrogen melt line up to pressures of 44~GPa have led to controversial 
conclusions regarding the existence of this maximum. Here we report 
{\it ab initio} calculations that establish the melt line up to 200~GPa. We
predict that subtle changes in the intermolecular
interactions lead to a decline of the melt line above 90~GPa. 
The implication is that as solid molecular hydrogen is compressed, 
it transforms into a low-temperature quantum fluid before becoming a monatomic 
crystal.
The emerging low-temperature phase diagram of hydrogen and its
isotopes bears analogies with the familiar phases of $^3$He and $^4$He,
the only known zero-temperature liquids,
but the long-range Coulombic interactions and the large component mass ratio
present in hydrogen would ensure dramatically different properties\cite{7,babaev}.
}

%
%

The possible existence of low-temperature liquid phases of compressed
hydrogen has been  rationalized with arguments based on  the nature of
effective pair  interactions and of the quantum dynamics at high
density, resulting in  proton-proton  correlations insufficient for
the stabilization of a  crystalline phase\cite{10}. But so far there
has been no conclusive  evidence establishing whether hydrogen
metallizes at low temperature as a solid  (the more widely  accepted
view  to date) or as a liquid. Measurements and theoretical
predictions of the near-ground state high-pressure phases\cite{mao} of
hydrogen  have proven to be difficult because of the light atomic
mass, significant quantum effects and  strong electron-ion
interactions. In this regard, the finite temperature liquid-solid
phase boundary predicted here is especially  valuable for
understanding the manner in which hydrogen metallizes.

The appearance of a  maximum melting temperature in hydrogen is in
itself a manifestation of an unusual physical phenomenon.  The
few systems with a negative melt slope involve either open crystalline
structures, such as water  and graphite, or in the case of closed packed
solids, a promotion of  valence electrons to higher orbitals upon
compression (6s to 5d in cesium\cite{9}, for example).  In these
cases, the liquid is denser than the solid when they coexist, possibly
because of structural or electronic transitions taking place
continuously  in the liquid, as a function of pressure,  but only at
discrete pressure  intervals in the solid.

In contrast, recent experiments\cite{3} have shown that hydrogen phase I 
-- a solid
structure with rotationally-free molecules associated with the sites
of a hexagonal closed packed (hcp) lattice -- persists below the
liquid transition up to at least 150 GPa, i.e. well beyond the melt
curve maximum predicted here.  The promotion of electrons to higher
orbitals in hydrogen can also be ruled out  because of the prohibitive
high energy involved. Alternatively, it has been suggested that
dissociation in the fluid, either gradual\cite{2}  or following a
first-order liquid-liquid (LL) phase transition\cite{11,12},  may be
the origin of a maximum in the melt line. This idea 
is not supported by our results. Instead, we explain the
physical  origin of the maximum in terms of changes in the  {\em
intermolecular} interactions -- a mechanism  significantly different
from that expected from familiar phenomenological models.

\begin{figure}[b]
  \includegraphics[width=0.45\textwidth,clip]{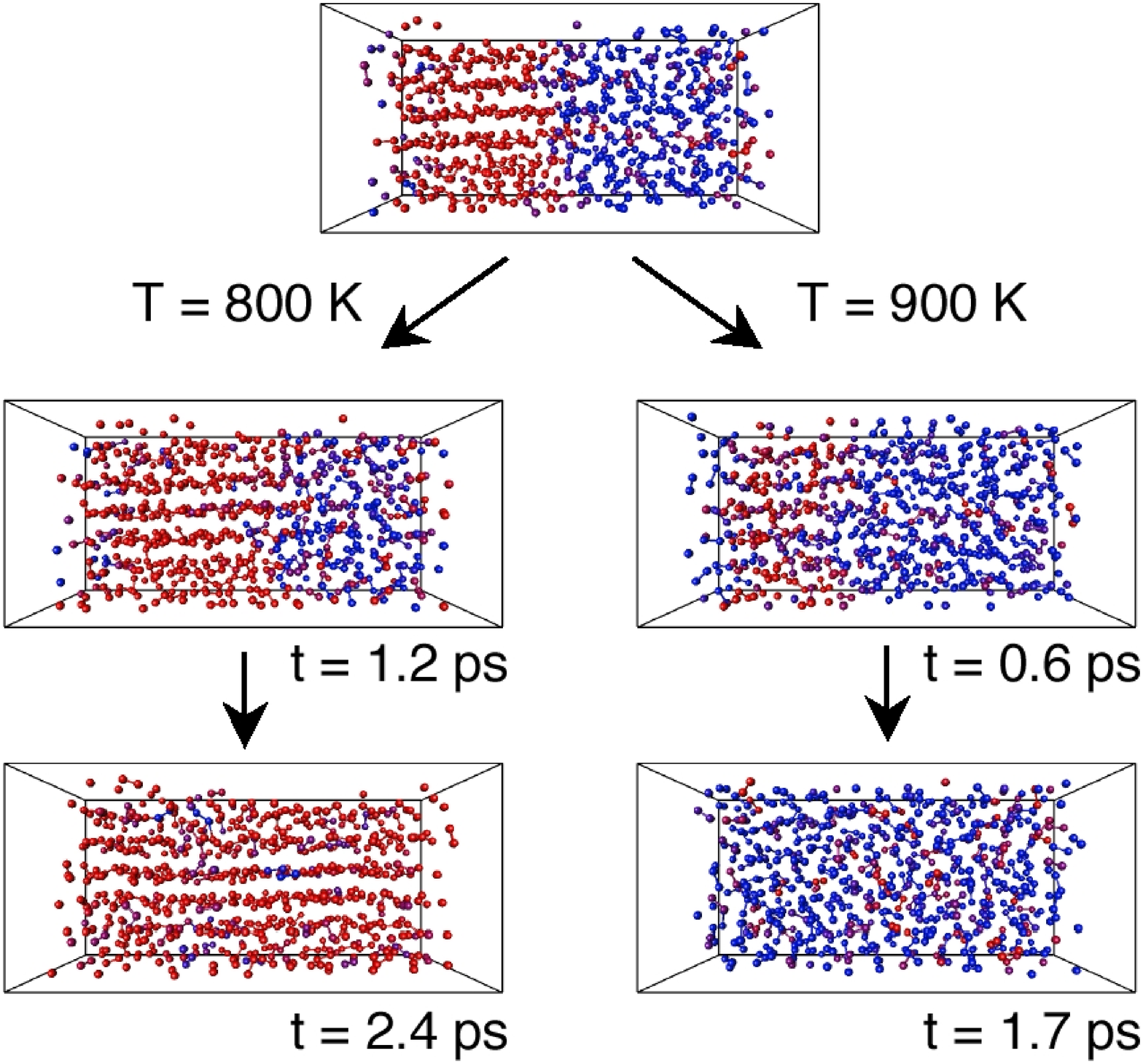}
  \caption{\label{fig_snapshots}Snapshots from two-phase MD
  simulations at $P= 130$~GPa and temperatures below and above the
  melting temperature. A quantitative assessment of the instantaneous
  local order environment around each molecule has been performed as
  described in Ref.~\onlinecite{17}, and compared with  single-phase
  solid and liquid simulations.  Molecules are coloured according to
  the arrangement of their nearest neighbours, red and blue
  representing configurations uniquely characteristic of the hcp solid
  and liquid at the given $P$ and $T$, respectively  (the $Q_4$ order
  parameter\cite{17} has been used for the colour map).  The systems
  reach  equilibrium at the target average temperatures (800 and 900
  K) over a time interval of $\sim 0.5$ ps, which is followed by
  periods of coexistence lasting from 1 to several picoseconds.   The
  phase transitions are observed by monitoring changes in the
  diffusion constants, pair  correlation functions, specific volumes
  and local order parameters.  In addition, some noticeable
  correlations between temperature and diffusion variations indicate
  the exchange of latent heat.   }
\end{figure}

The approach taken here to compute the melt line is one of direct simulation 
of the melting process. 
Hysteresis effects of super-heating or super-cooling during
the phase transition are avoided by simulating  solid and liquid
phases in coexistence. The validity of this method is then assessed
by reducing size effects in such a way that  realistic thermodynamic
processes can be mimicked. This technique, known as two-phase
simulation, has been mostly applied with model potentials\cite{14},
thus allowing the simulation of large systems,  and only recently it
was demonstrated\cite{15,16} that implementations  within the
framework of first principles molecular dynamics (MD)  are feasible.

Two-phase simulations (Fig.~1) are performed for various pressure
and temperature conditions to predict the melt curve of
hydrogen up to 200~GPa  (Fig.~2). As discussed in Refs.~\onlinecite{2} and 
\onlinecite{3}, the
experimental  data available up to 44~GPa  can be equally  well fitted
with several empirical melt  equations, leading to qualitatively  different
conclusions regarding the further rise or fall of the phase boundary.
Our results at 50~GPa compare favourably with recent
measurement\cite{3}, and the  calculations at 130  and 200 GPa predict
a negative slope in this pressure  region.

\begin{figure}[b]
  \includegraphics[width=0.45\textwidth,clip]{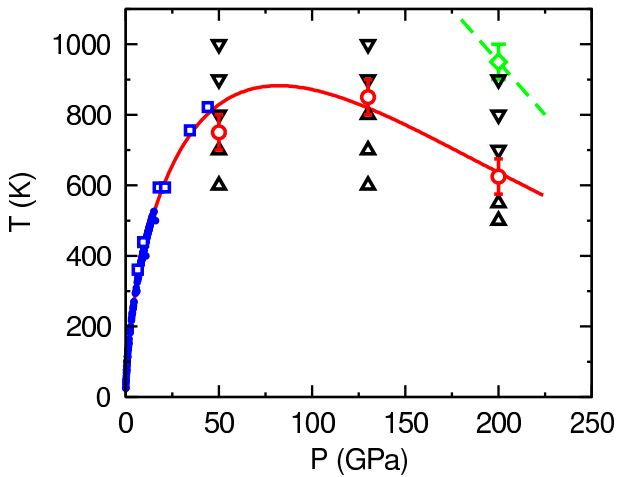}
  \caption{\label{fig_melt_curve}Melt curve of hydrogen predicted from
  first principles MD. The filled circles are experimental data from
  Refs.~\onlinecite{1} and \onlinecite{2} and references therein, 
  and the open squares
  are measurements  from Ref.~\onlinecite{3}.   Triangles indicate
  two-phase simulations where solidification (up) or melting (down)
  have been observed, and bracketed melting temperatures  ($T_m$) are
  represented by open circles. As the phase boundary is approached,
  the period of coexistence increases and eventually the outcome
  becomes dependent on the choice of simulation parameters. This
  degree of arbitrariness is reflected in the error bars of $T_m$,
  which also include the standard deviation of the temperatures
  collected  during the MD simulations.  All experimental and
  theoretical points are given equal weight and fitted with a Kechin
  melt equation\cite{18}  (solid line in the figure): $T_m =
  14.025\left(1+P/a\right)^b\exp(-cP)$ K,  where $P$ is in units of
  GPa, $a=0.030355$, $b=0.59991$, and $c=0.0072997$. The open diamond
  marks the liquid-liquid transition from molecular to non-molecular fluid at
  200 GPa, and the estimated slope of this phase boundary is given by
  the dashed line. The error bar on the  diamond symbol indicates the
  hysteresis effects during the simulation of the liquid-liquid transition.  
 }
\end{figure}

Previous first-principles calculations\cite{11,19} indicated the
existence of a sharp transition from molecular to non-molecular fluid, thus
raising the possibility that the melt curve maximum is  a
liquid-liquid-solid triple point. We therefore
studied the properties of the liquid at  temperatures above
melting. In our simulations, we do find a
first-order liquid-liquid phase transition from molecular to dissociating 
fluid,
but this occurs at temperatures higher than  those of the melting
line;  at 200 GPa, the liquid-liquid transition takes place between 900 and
1,000~K (Fig.~2). Thus, in the entire pressure range considered here,
the liquid   remains molecular below 900~K, with molecules stable over
the simulation times of several picoseconds. We emphasize that on heating 
above a critical temperature, the transition to a
non-molecular fluid takes place very rapidly, over a couple of hundred
femtoseconds of simulation time.  When the fluid is subsequently
quenched, the molecules recombine and the  system reverts to the
molecular state. Therefore, we conclude  that the change from a
positive to a negative melt line slope happens gradually and we
emphasize that it is not directly related to molecular  dissociation.
However, extrapolations of the melt line and the liquid-liquid phase
transition indicate a triple point at $\sim 300$~GPa and
400~K. Above this pressure, the solid is expected to melt into a 
metallic liquid.

\begin{figure}[b]
  \includegraphics[width=0.45\textwidth,clip]{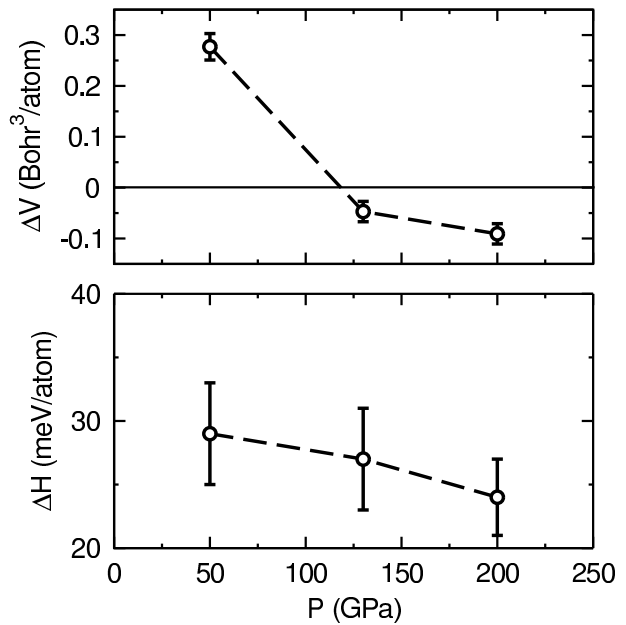}
  \caption{\label{fig_dv_dh}Difference of the specific volumes
  ($\Delta V$) and  enthalpies ($\Delta H$) between the liquid and
  solid phases at the melting temperatures determined from the
  two-phase simulations. The reported uncertainties include standard
  deviations collected during the MD simulations and observed  size
  effects from singe-phase simulations with 360 and 768 atom
  supercells.  The data are used to compute melt slopes from the
  Clausius-Clapeyron equation: $dT_m/dP = T_m \Delta V/\Delta H$; the
  values obtained for 50, 130 and 200 GPa are ($6.5\pm 1.2$),
  ($-1.4\pm 0.6$)  and  ($-2.3\pm 0.6$) K/GPa, respectively. For
  comparison, the slopes from the melt line fit in Figure 2 are 3.9,
  -2.2 and -2.7 K/GPa, but we note that  these are weighted heavily by
  the experimental data, especially at lower  pressure. 
 }
\end{figure}

To estimate the errors originating from
finite system size effects, we have computed the  ground state
pressure and energy for different size simulation cells sampled at the
the $\Gamma$-point and with a 2x2x2 {\bf k}-point mesh.  The pressure
and energy {\em differences}  between the solid and the liquid are
converged to  within $0.3$~GPa and~$1$ meV per ion, and the estimated
errors are included in Fig.~3.  A question that remains to be
answered is whether the size of the system considered here is
sufficient to simulate liquid-solid coexistence.  To address this, 
we carried out MD simulations of single-phase solid  and
liquid hydrogen at the melting temperatures determined in the
two-phase simulations. The differences in the specific volumes and
enthalpies obtained in this way (Fig.~3) are then used to compute
the melt slopes using the Clausius-Clapeyron equation. The agreement
with the results from two-phase simulations gives us confidence that
the computed melt curve is thermodynamically consistent.

We also discuss the validity of the principal
approximations in our  theoretical method: (1) adiabatic
interactions, (2) the generalized gradient approximation (GGA) for the
exchange-correlation energy, and (3) classical treatment of ion
motion.  Electron-phonon interactions are significant in the solid
phase when the direct band gap is comparable with the phonon
frequencies; an event that  eventually occurs at high densities. But
at 200~GPa and 600~K -- the conditions examined here where the
electron-phonon coupling is expected to be most pronounced -- the GGA
band gap is still $\sim 2$~eV.  We note that the GGA tends to
systematically underestimate band gaps; this has been shown for
various materials, including hydrogen\cite{czl92}.  However, because
the highest vibrational frequencies in hydrogen are $\sim 0.5$~eV, the
computed GGA band gap -- a lower bound to the actual value -- is
already sufficient to establish that non-adiabatic effects are indeed
negligible.

Local density approximations are also expected to favour the phase
with more delocalised electrons; previous studies have shown a
consistent tendency of the GGA to underestimate melting temperatures,
such as those of lithium hydride\cite{15} and aluminium\cite{16}.  
Here, this effect is expected to be very small because the
molecular bonding properties and the atomic coordination differ little
in  the solid and liquid phases. This also implies that the zero-point
energy bound in the vibron motion is similar in the two
phases. Indeed, velocity-velocity autocorrelation analysis carried out
on our MD trajectories shows that vibron frequency distributions
(containing all vibrational modes) differ by less than 10~cm$^{-1}$ in
the solid and the liquid between 50 and 200 GPa. In addition, we have
computed the first-order quantum correction to the ionic free
energies. It is given by the Wigner-Kirkwood formula,  $\Delta F =
\hbar^2/(24 k_B^2T^2) \sum_i \langle \mathbf{F}_i^2 \rangle /m_i$,
where the average is over the classical ensemble, and $\mathbf{F}_i$
and $m_i$ are the ionic forces and masses. The {\em difference} in the
quantum corrections for the two phases obtained in this way at
pressures of 50, 130 and 200 GPa remains less than 2 meV for
temperatures near the computed melt curve.

We now turn our attention to the physical origin of the maximum in the
melt curve of hydrogen. The transition from an ordered to a disordered
state is governed by the relative importance of entropy and energy. At
high pressure, the interactions become mostly repulsive, and it is
reasonable to expect that a  deviation from a crystalline arrangement
becomes progressively more unfavourable in terms of enthalpy. This is
indeed the case for most materials, as is reflected in the positive slopes
of their melt curves. On the other hand, attractive many-body
interactions can soften the steep increase of the repulsive  forces at
high density\cite{hmf90,nbb03}. Although, in principle, such a softening opens
up the possibility for a melt curve maximum\cite{yoshida}, it does not
appear to be a sufficient condition for its existence. Indeed,
softening of the repulsive part of the pair potentials has been
observed for a number of materials for which there is no experimental
indication of a melt line turnover. As will be shown below, the
physics behind the unique melt curve of hydrogen is related to the
{\em different rate} at which the repulsive interactions (at
intermolecular range) are weakened in the solid and liquid phases as a
function of pressure.

\begin{figure}[t]
  \includegraphics[width=0.45\textwidth,clip]{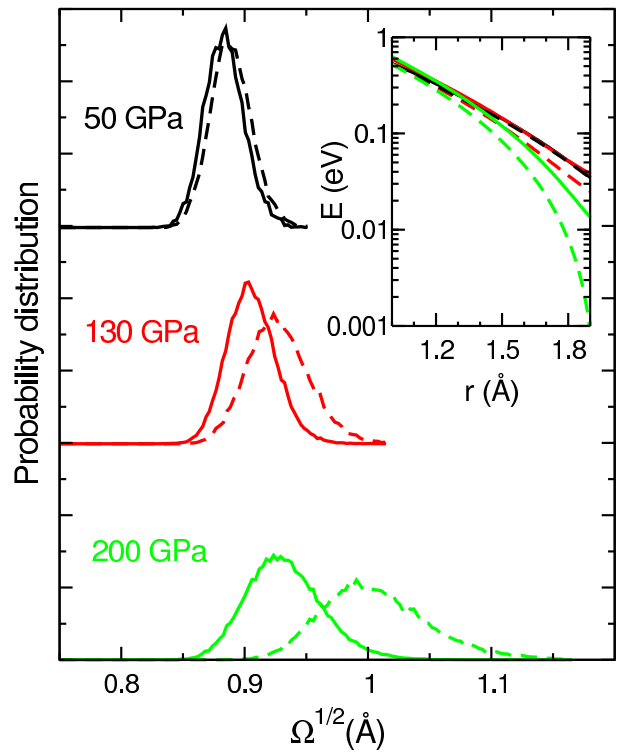}
  \caption{\label{fig_mlwf} 
The probability distribution of MLWF
spreads \cite{marzari} at $T$=700 K and $P$=50 (black curves), 130 (red
curves) and 200 GPa (green curves). The solid and dashed curves correspond
to distributions collected from the solid and liquid phases, respectively.
The inset shows the effective hydrogen-hydrogen potentials obtained directly
from the simulation trajectories with the force-matching procedure described
in Ref.~\onlinecite{ipb04} (the same coloring has been used). 
As the pressure is
increased, the spreads of the MLWF are shown to increase faster in the
liquid than in the solid phase. In addition, effective
potentials soften more quickly in the liquid than in the solid phase at the 
nearest neighbor
molecular separation distance. Note that the potentials at 50 GPa in the 
liquid and solid phases, and at 130 GPa in the solid phase, are
indistinguishable on this scale. 
 }
\end{figure}

To investigate how many-body interactions give rise to
repulsive force softening in the liquid and the solid, we have
performed an analysis based on maximally-localized Wannier functions
(MLWF). These functions have been shown\cite{marzari} to  provide an
intuitive description of chemical bonds in condensed phases and they
have recently been used\cite{souza} to investigate the infrared
activity of candidate structures of hydrogen phase III. We have
computed MLWF along the 700~K isotherm, at various pressures, and find
that their  spreads (Fig.~4)  increase faster in the fluid than in
the solid, as the system is compressed. Importantly, the differences in
the two phases come from the tails of the MLWF, while their profiles
around the centres of the molecular bonds remain nearly the
same. Therefore, the large MLWF spread observed in the high-pressure fluid
comes from  the increased overlap of
the molecular orbitals, here represented by the MLWF. Because the
specific volumes of the solid and the liquid are very  similar under
the conditions examined here, the physical origin of the larger spread
in the fluid can be attributed to the disorder in the liquid phase.  
Furthermore, the
tails of the MLWF in  the fluid are characterized by an asymmetry,
which can be viewed as an effective intermolecular charge transfer,
giving rise to an  enhanced infrared activity in the liquid
phase.

The properties of the MLWF found in our calculations are indicative
of the behaviour of repulsive interactions in the solid and liquid as
a function of pressure. Effective hydrogen-hydrogen pair potentials
derived from our {\it ab initio} MD trajectories using a force
matching technique\cite{ipb04} are displayed in the inset of Fig.~4.
As in the MLWF behaviour, we observe a stronger softening of the
potential in the liquid than in the solid. Our
Wannier function analysis indicates that this increased softening 
is the result of charge transfer processes, enhanced by
disorder, which appear to have a qualitative character similar to
those proposed in Ref.~\onlinecite{hemley} for phase III of
solid hydrogen.  We also note that the attenuation of repulsive
interactions observed here in the solid phase correlates well with the
analysis of Raman spectroscopic data\cite{silvera2}, 
which has indicated softening
of the effective intermolecular force constants above $\sim$100 GPa.

To summarize, we have predicted the melt curve of hydrogen between 50
and  200~GPa from first principles. Above approximately 82~GPa, the
melt line has a negative slope -- a phenomenon that we have related to
softening of the intermolecular interactions, which occur at a faster
rate in the liquid than in the solid, as a function of pressure. Our
results provide strong evidence that above $\sim$300~GPa the solid
melts into a metallic liquid and  that a low-temperature metallic
quantum fluid will exist at pressures near 400~GPa.  Although we have
only directly probed temperatures as low as 600~K, our conclusions are
likely to hold at lower temperatures as well. Indeed, quantum ion
motion usually favours phases with lower ion-ion correlations, which
would further stabilize the low-temperature liquid phase found
here. In addition, our findings are consistent with pressure estimates
from experiments where hydrogen is  expected to
metallize\cite{mao,lot}. Finally, we note that the observed increase
in MLWF spreads when going from the high-pressure solid to the
molecular liquid phase points at the presence of dynamically-induced
intermolecular charge transfer  (dipole moments); this indicates that
the transition from solid to liquid at high pressure is
accompanied by an increase of the hydrogen infrared
absorption. Therefore, we propose that measurements of infrared
activity could be used to verify experimentally the predicted turnover
of the melt curve.

\vspace{0.6cm}
{\bf \large \noindent Methods}

\vspace{0.2cm}
{\bf \noindent Molecular dynamics}\\[0.1cm]
\noindent Our {\it ab initio} MD simulations are carried out with the 
GP code, written by F. Gygi,
and using the Car-Parrinello (CP) method\cite{20}; the many-body
electron problem was solved quantum-mechanically within the generalized
gradient approximation\cite{21} of density functional theory,  while
the ions are propagated classically (the validity of this
approximation for the problem at hand  is discussed in the text). The
actual calculations are performed for deuterium instead of
hydrogen. The reason for this is that  in the classical-ion picture,
for the quantities of relevance here there is no distinction between
hydrogen and its isotopes, but the substitution with heavier ions
allows us to integrate  the equations of  motion with reasonably large
time steps:  2 and 3~a.u.  (1~a.u. = 0.0242~fs) depending on the
density.  Furthermore, we use a $\Gamma$-point sampling of the
Brillouin zone,  Troullier-Martins pseudopotentials, and 60~Ry energy
cut-off for the plane wave expansion of the Kohn-Sham orbitals. The
accuracy and transferability of our pseudopotential has 
previously been extensively tested\cite{19}. We have explicitly verified
that the electrons remain near the Born-Oppenheimer surface throughout
the simulations by comparing pressures determined in the CP runs with
those determined by fully optimizing wavefunctions for the same ionic
trajectories. The use of CP-MD results in a systematic shift of
about 0.5 GPa both in the solid and liquid phases.

\vspace{0.2cm}
{\bf \noindent Two-phase simulations}\\[0.1cm]
\noindent We perform the two-phase simulations 
by equilibrating solid (in the hcp structure) and liquid
phases  separately in 360-atom supercells with fixed geometry and
applied periodic boundary conditions.  The atomic arrangements
obtained in this way  are  then merged, and the liquid and solid
parts of the resulting 720-atom  supercells are also heated
independently to reduce the  strain  near the newly-formed
interface. The whole equilibration process  typically lasts for several
picoseconds of simulation time. The MD runs of  coexisting phases are
then performed in the $N-P-T$ ensemble   (constant number of particles,
pressure and temperature respectively).  The actual system being
simulated initially represents  semi-infinite alternating slabs (by
applying periodic boundary conditions) of solid and liquid, but by the 
end of the runs only one
of the two phases, the one that is stable at the chosen $P$ and
$T$, fills the entire volume.

%
%


{\small
{\bf\noindent Acknowledgments} 
We gratefully acknowledge S. Izvekov for providing the force-matching
routines that were used in analysis shown in the inset of Figure 4, 
and F. Gygi for many useful discussions.
This work was performed under the auspices of the U.S. Dept. of  Energy at 
the University of California/Lawrence Livermore National Laboratory.
\\[0.5cm]
{\bf Competing interests statement} The authors declare that they have
no competing financial interests.
\\[0.5cm]
{\bf Correspondence} and requests for materials should be addressed 
to S.A.B. (bonev1@llnl.gov).
}
\\[0.5cm]

\end{document}